\documentclass[fleqn,10pt]{wlscirep}
\usepackage[utf8]{inputenc}
\usepackage[T1]{fontenc}

\usepackage{amsmath}
\usepackage{amssymb}
\usepackage{graphicx}
\usepackage{mathrsfs}
\usepackage{enumerate}
\usepackage{tikz}
\usepackage{pgfplots}
\usepackage{placeins}

\renewcommand{\ge}{\geqslant}

\newcommand{\vell}{\boldsymbol\ell}

\newcommand{\cB}{\mathcal{B}}
\newcommand{\cL}{\mathcal{L}}
\newcommand{\cS}{\mathcal{S}}

\newcommand{\EE}{\mathbb{E}}

\title{Beyond Dunbar circles: a continuous description of social relationships and resource allocation}

\author[1,2]{Ignacio Tamarit}
\author[1,2,3,4]{Angel S\'anchez}
\author[1,2,3,4,*]{Jos\'e A. Cuesta}

\affil[1]{Grupo Interdisciplinar de Sistemas Complejos, Departamento de Matem\'{a}ticas, Universidad Carlos III de Madrid, 28911 Legan\'es, Madrid, Spain.}
\affil[2]{Unidad Mixta Interdisciplinar de Comportamiento y Unidad Social (UMICSS) UC3M-UV-UZ, Universidad Carlos III de Madrid, 28911 Legan\'es, Spain.}
\affil[3]{Institute for Biocomputation and Physics of Complex Systems (BIFI), University of Zaragoza, 50018 Zaragoza, Spain.}
\affil[4]{UC3M-Santander Big Data Institute (IBiDat), 28903 Getafe, Spain.}

\affil[*]{cuesta@math.uc3m.es}


\begin{abstract}
We discuss the structure of human relationship patterns in terms of a new formalism that allows to study resource allocation problems where the cost of the resource may take continuous values. This is in contrast with the main focus of previous studies where relationships were classified in a few, discrete layers (known as Dunbar's circles) with the cost being the same within each layer. We show that with our continuum approach we can identify a parameter $\eta$ that is the equivalent of the ratio of relationships between adjacent circles in the discrete case, with a value $\eta\sim 6$. We confirm this prediction using three different datasets coming from phone records, face-to-face contacts, and interactions in Facebook. As the sample size increases, the distributions of estimated parameters smooth around the predicted value of $\eta$. The existence of a characteristic value of the parameter at the population level indicates that the model is capturing a seemingly universal feature on how humans manage relationships. Our analyses also confirm earlier results showing the existence of social signatures arising from having to allocate finite resources into different relationships, and that the structure of online personal networks mirrors those in the off-line world.
\end{abstract}

\pgfplotsset{compat=1.16}
\begin{document}

\flushbottom
\maketitle

\thispagestyle{empty}

\section{Introduction}

Human relationships show clear organizational patterns. Numerous studies reveal that we structure our personal relationships into groups (also known as circles) whose inclusive sizes follow an approximately geometric progression with a scale factor of 3:\cite{sutcliffe2012relationships,dunbar2014social,hill2003social,dunbar2015structure,zhou2005discrete,robinprsa2020} 5, 15, 50, and 150 (evidence for yet another circle of size approximately 500 has been provided recently\cite{joseluis2019}). These circles display marked differences regarding emotional closeness and time devoted to relationships, which correlate with one another.\cite{oswald2004friendship,roberts2009exploring,pollet2013going} Of all the time we dedicate to our social life, approximately $40\%$ is devoted to people in our most intimate circle (support clique), $20\%$ to close relationships (sympathy group), and the remaining $40\%$ to the rest of relationships\cite{sutcliffe2012relationships}---progressively devoting less to those more distant. In addition, maintaining social relationships is not only costly from the perspective of the time they require, but it is also costly in cognitive terms. Studies combining neuroimaging techniques and cognitively demanding tasks  show that individual differences in the volume of the orbitofrontal cortex (a specific region of the neocortex) explained differences in mentalising skills, and those, in turn, were able to explain differences in network size.\cite{powell2012orbital} 


The connection between the hierarchical structure of personal networks and the costs associated to maintaining them has recently been formalized as a simple model of resource allocation.\cite{tamarit:2018} This model is based on two strong and robust empirical observations. Firstly, the number of relationships a person has, $\cL$, tends to be stable over time.\cite{saramaki2014persistence} Secondly, there are different \emph{costs} to maintaining different types of relationships,\cite{oswald2004friendship,miritello2013time,dunbar2018anatomy} and the total cognitive resources we apply to them, $\cS$, are limited. The maximum entropy principle\cite{jaynes2003probability} is then used to add that information to a multinomial prior, and the result is a posterior distribution that measures the likelihood of different allocations of resources to relationships characterised with different costs (see Ref.~\citeonline{tamarit:2018} for details). Generally, this distribution agrees with the organization of relationships in circles as empirically observed. 
The comparison of the model introduced in Ref.~\citeonline{tamarit:2018} to the available data on real social systems requires treating relationships as a set of $r$ discrete categories (layers) that are defined based on their intensity. Circles are then defined as the union of layers up to a given intensity. This approach is particularly convenient when intensity is measured in Likert scales, as it is often the case when data is obtained via questionnaires. 

There are, however, alternative ways of measuring tie strength which do not rely on a discrete scale like the one used with the circles. Good examples are frequency of contact,\cite{roberts2009exploring} time spent together,\cite{mastrandrea2015contact} or number of messages (information) exchanged.\cite{saramaki2014persistence,dunbar2015structure} Even though some of these quantities could be technically regarded as discrete, the fact that they consist of hugely many possibilities makes this viewpoint rather impractical. More importantly, these measures do not have clear upper and lower bounds (what is the shortest duration of a call to be considered a contact?) that play the role of first and last layers, respectively. This calls for a more general version of the model that would allow us to consider intensities of high granularity, possibly continuum. On the other hand, such a model would be conceptually very general in so far as many resource allocation problems are of a continuous or quasi-continuous nature.



The purpose of this paper is therefore to introduce a general model in which the allocated amount of resource can take any positive real number.  After going through the description of the model and its mathematical study, we apply it to three different datasets in which the intensity of personal relationships is measured with continuous variables: face to face contact time,\cite{Isella2011crowd} number of messages between Facebook users,\cite{arnaboldi2012analysis} and number of phone calls exchanged.\cite{saramaki2014persistence} Our analyses unveil the existence of a structure similar to that found when intensities are considered as discrete categories, thus showing that there is no need to exogenously categorize the data to understand its structure. More importantly, we prove the existence of a new universal scale parameter $\eta$, which replaces (and is consistent with) the scale factor $\sim3$ ubiquitously found in the discrete scenario with social relationships.

\section{Model description}
\label{sec:model}

We introduce our model from a completely abstract viewpoint, by starting from an individual that must distribute a limited amount of some resource among an assortment of $N$ different choices. We will denote by $\cL$ the average number of different choices the individual makes and by $\cS$ the average amount of resource invested in them (irrespective of which magnitude we use to measure it). In the particular example of the ego-networks that we will explore in more depth here, $\cL$ represents links to alters and $\cS$ the individual's cognitive resources devoted to keep those links. 
At this point, however, we are not concerned by the precise nature of these two magnitudes, only with their existence and their limited values.

For the time being, let us assume that all possible choices can be classified within $r$ different categories, each of them bearing a different cost (in terms of resource invested) $s_{\rm max}=s_1>s_2>\cdots>s_r=s_{\rm min}$. A maximum entropy analysis shows that the probability that an individual chooses $\ell_k$ elements within the category $k$ ($k=1,\dots,r$) is given by \cite{tamarit:2018}
\begin{equation}
P(\ell_1,\dots,\ell_r|\cS,\cL,N)=\sum_{L=0}^NP(\ell_1,\dots,\ell_r|\sigma,L)\cB(L,\cL/N,N),
\qquad \sigma\equiv\frac{\cS}{\cL},
\end{equation}
where $\mathscr{B}(L,\cL/N,N)$ is the binomial distribution for the total number of choices, and
\begin{equation}
P(\ell_1,\dots,\ell_r|\sigma,L)=\frac{L!}{\ell_1!\cdots\ell_r!}
\frac{e^{-\hat{\mu}\sum_{k=1}^rs_k\ell_k}}{\left(\sum_{k=1}^re^{-\hat{\mu} s_k}\right)^L}
\delta\left(L,\sum_{k=1}^r\ell_k\right).
\label{eq:Pell-sigma}
\end{equation}
Here $\delta(x,y)=1$ if $x=y$ and $0$ otherwise, and the parameter $\hat\mu=\hat\mu(\sigma)$ is determined by the equation
\begin{equation}
\sigma=\frac{\sum_{k=1}^rs_ke^{-\hat\mu s_k}}{\sum_{k=1}^re^{-\hat\mu s_k}}.
\label{eq:sigma-mu}
\end{equation}

The cost is the only variable that distinguishes different choices, so in order to make discrete categories it is natural to split the whole range of costs uniformly. Thus,
\begin{equation}
s_k=s_{\rm max}-(s_{\rm max}-s_{\rm min})\frac{k-1}{r-1}, \qquad k=1,\dots r,
\label{eq:linearcosts}
\end{equation}
with $k=1$ ($k=r$) corresponding to the most (least) costly category, following the standard convention used in previous studies. Substituting this form for $s_k$ into the probability distribution \eqref{eq:Pell-sigma} we obtain
\begin{equation}
P(\ell_1,\dots,\ell_r|\sigma,L)=\left(\frac{e^{\mu}-1}{e^{\mu r}-1}\right)^L
\frac{L!}{\ell_1!\cdots\ell_r!}e^{\mu L_1}\delta\left(L,\sum_{k=1}^r\ell_k\right),
\qquad L_1\equiv\sum_{k=1}^r(k-1)\ell_k,
\label{eq:Pell-sigma-lin}
\end{equation}
with $\mu\equiv\hat{\mu}(s_{max}-s_{min})/(r-1)$.

This is nothing but the probability distribution of links in an ego-network that was obtained in Ref.~\citeonline{tamarit:2018}, but our goal is to describe a continuum of levels, not these discrete categories. To that purpose we need to take the limit $r\to\infty$ appropriately: levels will now be described by a continuous index $t\equiv(k-1)/(r-1)\in[0,1]$. Notice that $t=0$ corresponds to $s_{max}$ and $t=1$ to $s_{min}$, so the parameter $t$ can also be interpreted as a sort of `distance' to the corresponding choice. As a matter of fact, it is possible to parameterize everything in terms of cost rather than distance, by introducing $s=1-t$. 

In the same limit of infinitely many categories, $r\ell_k\to\ell(t)$, so that $\ell_k$ gets transformed into a density of links $\ell(t)\,dt$. Accordingly,
\begin{equation}
L\to\lim_{r\to\infty}\sum_{k=1}^r\ell_k=\int_0^1\ell(t)\,dt\equiv\tilde{L},
\qquad
\mu L_1\to\mu\lim_{r\to\infty}\sum_{k=0}^{r-1}k\ell_{k+1}=\eta\int_0^1t\ell(t)\,dt
\equiv\eta\tilde{L}_{1},
\label{eq:LL1limits}
\end{equation}
where $\eta\equiv\mu(r-1)=\hat{\mu}(s_{max}-s_{min})$. Furthermore, when the number of layers $r$ is large, the probability that two individuals belong to the same category goes to zero, so $\ell_k$ should be either $0$ or $1$ for all $k$. This implies that $\binom{L}{\vell}\to L!$. 

In order to proceed now with the distribution of links \eqref{eq:Pell-sigma-lin}, it has to be realized that in this limit it becomes quite an unmanageable object---a path integral. There are two ways to circumvent this technical problem. The first one amounts to calculating the limit of averages. For the second one we should realise that, in the limit $r\to\infty$, the only dependence on $\ell(t)$ is through the moment $L_1$, so instead of dealing with a limit of \eqref{eq:Pell-sigma-lin} it is better to take the limit of the probability distribution
\begin{equation}
P(L_1|\sigma,L)=\sum_{\ell_1,\dots,\ell_r\ge 0}P(\ell_1,\dots,\ell_r|\sigma,L)
\delta\left(L_1,\sum_{k=1}^r(k-1)\ell_k\right)=W(L,L_1)
\left(\frac{e^{\mu}-1}{e^{\mu r}-1}\right)^Le^{\mu L_1},
\label{eq:PL1}
\end{equation}
where $W(L,L_1)$ is a factor that only depends on $L$ and $L_1$ and whose specific form does not concern us at this point. The first approach will be useful to obtain the expected distribution of choices as a function of their costs; with the second one we will derive a Bayesian estimate of the parameter $\mu(\sigma)$ in the limit $r\to\infty$.

\subsection{A continuum version}

\begin{figure}
    \centering
    \includegraphics[width=80mm]{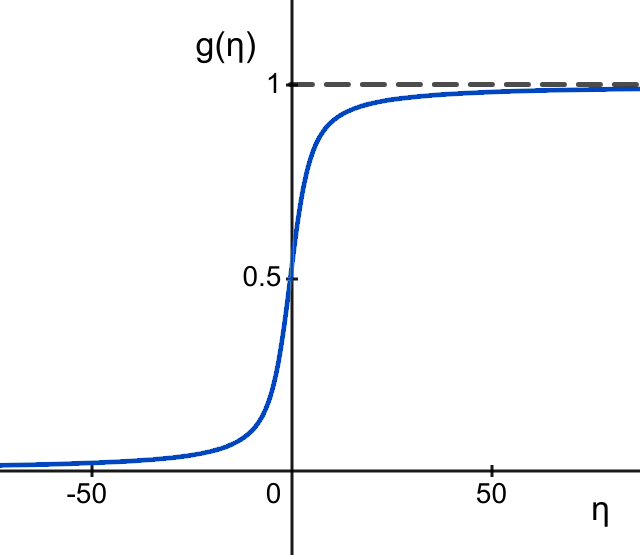}
    \caption{Plot of the function $g(\eta)$ that determines the cost associated to a value of the scaling parameter $\eta$ through equation~\eqref{eq:geta}.}
    \label{fig:g-eta}
\end{figure}

Using the distribution \eqref{eq:Pell-sigma-lin} it is possible to calculate $\epsilon_k$, the expected number of choices from category $k$, as well as $\chi_k$, the expected number of choices with costs larger than or equal to that of category $k$. The latter is what in the literature of ego-networks is referred to as a social ``circles''.\cite{mccarty2019} It is straightforward to obtain the expression\cite{tamarit:2018}
\begin{equation}
\epsilon_k\equiv\frac{\EE(\ell_k)}{L}=\frac{e^{k\mu}-e^{(k-1)\mu}}{e^{r\mu}-1},
\qquad \chi_k\equiv\sum_{j=1}^k\epsilon_k=\frac{e^{k\mu}-1}{e^{r\mu}-1}.
\label{eq:disc-circles}
\end{equation}
Taking the limit $r\to\infty$ transforms these expected values into their continuous counterparts:
\begin{equation}
\epsilon_k\to\epsilon(t)\,dt=\frac{\eta e^{\eta t}}{e^{\eta}-1}\,dt,
\qquad \chi_k\to\chi(t)=\frac{e^{\eta t}-1}{e^{\eta}-1}=\int_0^t\epsilon(u)\,du.
\label{eq:chit}
\end{equation}
In particular, $\chi(t)$ is the fraction of links whose ``distance'' to the individual is not larger than $t$. Finally, we must find the relationship between the continuum parameter $\eta$ and the discrete parameter $\sigma$. In order to do that, we start off from equation \eqref{eq:sigma-mu} which, after substituting \eqref{eq:linearcosts}, becomes
\begin{equation}
\frac{s_{max}-\sigma}{s_{max}-s_{min}}
= e^{\mu}\frac{(r-1)e^{r\mu}-re^{(r-1)\mu}+1}{(r-1)(e^{r\mu}-1)(e^{\mu}-1)}.
\end{equation}
The continuum limit ($r\to\infty$, $\mu\to 0$ with $\eta=\mu(r-1)=\text{constant}$) of this expression yields
\begin{equation}
t=\frac{s_{max}-\sigma}{s_{max}-s_{min}}=\frac{e^{\eta}}{e^{\eta}-1}-\frac{1}{\eta}\equiv g(\eta),
\label{eq:geta}
\end{equation}
an implicit equation whose solution provides the sought for dependence $\eta=\eta(\sigma)$. Notice that 
\begin{equation}
\lim_{\eta\to-\infty}g(\eta)=0, \qquad \lim_{\eta\to 0}g(\eta)=\frac{1}{2},
\qquad \lim_{\eta\to +\infty}g(\eta)=1,
\end{equation}
and since $g'(\eta)>0$ for all $\eta \in \mathbb{R}$, equation~\eqref{eq:geta} has a unique solution for any $0<t<1$. As a matter of fact, $\eta=0$ for $t=1/2$, whereas $\eta>0$ for $t>1/2$ and $\eta<0$ for $t<1/2$---hence $\eta'(t)>0$ (see the plot of $g(\eta)$ in Figure~\ref{fig:g-eta}).

\subsection{Connection with the theory of ego-networks}
\label{sec:connection}

With the calculations above, we are now in a position to obtain a quantitative estimate of the parameter $\eta$ that determines the distribution in the continuum, for the specific application to Dunbar's social circles in this limit. Recall that in the social circles interpretation the choices are links to alters of an ego, cost means cognitive cost, and the categories describe layers of emotional closeness of the corresponding relationships.

For large values of $\mu$, equation~\eqref{eq:disc-circles} behaves as
\begin{equation}
\frac{\chi_{k+1}}{\chi_k}\sim
\begin{cases}
e^{\mu}, & \mu\to\infty, \\
1, & \mu\to -\infty.
\end{cases}
\label{eq:asympdisc}
\end{equation}
This shows, on the one hand, that in the ordinary regime ($\mu>0$) the circles (quantified by $\chi_k$) satisfy an approximate scaling relation, and on the other hand, that in the so-called ``inverse'' regime ($\mu<0$) the closest circle becomes overpopulated. Both behaviours have been properly documented in the literature.\cite{dunbar2015structure,zhou2005discrete,tamarit:2018,dunbar2018,lubbers2019}

The corresponding analysis for the continuum model requires that we determine the asymptotic behaviour, for large $\eta$, of the logarithmic derivative of $\chi(t)$, namely
\begin{equation}
\frac{\dot{\chi}(t)}{\chi(t)}=\frac{\eta e^{\eta t}}{e^{\eta t}-1}\sim
\begin{cases}
\eta, & \eta\to\infty, \\
0, & \eta\to -\infty.
\end{cases}
\label{eq:asympcont}
\end{equation}
As the discrete version of the left-hand side is $(\chi_{k+1}-\chi_k)/\chi_k\Delta t$, a comparison between \eqref{eq:asympdisc} and \eqref{eq:asympcont} in the ordinary regime leads to $\eta\Delta t\approx e^{\mu}-1$. Since $\Delta t\approx (r-1)^{-1}$, we obtain the equivalence
\begin{equation}
\eta\approx(r-1)(e^{\mu}-1).
\label{eq:scaling}
\end{equation}
Equation~\eqref{eq:asympcont} reveals that $\eta$ is the true underlying scaling factor of the circles. Therefore, the equivalence just derived implies that the value of $\mu$ in the discrete model must depend on the total number of circles $r$. This fact has been overlooked in previous analysis of the original circles model because of the implicitly assumption that there are $r=4$ circles in the structure of ego-networks.\cite{dunbar2015structure} If we set $r=4$ in \eqref{eq:scaling} and input the empirical scaling observed in this model $e^{\mu} \approx 3$,\cite{dunbar2015structure} we conclude that the scaling to be expected in a continuous setting of social relationships must be $\eta\approx 6$. This is a concrete prediction of the continuum model that needs to be tested against actual data.

\section{Data analysis}

In this section we will explore how this continuous model compares to actual data. We will use three datasets for this comparison: phone calls,\cite{saramaki2014persistence} face-to-face contacts,\cite{Isella2011crowd} and interactions between Facebook users.\cite{arnaboldi2012analysis} But before that we need to develop a formalism to make the fits and determine their confidence intervals.

\subsection{Bayesian estimate of the scaling parameter}
\label{sec:bayesian_estimate}

Starting from \eqref{eq:PL1} and assuming a noninformative uniform prior for $\mu$, it follows that, up to a normalising constant,
\begin{equation}
P(\mu|L,L_1)\propto\left(\frac{e^{\mu}-1}{e^{\mu r}-1}\right)^Le^{\mu L_1}.
\end{equation}
In the continuum limit and using the definitions~\eqref{eq:LL1limits},
\begin{equation}
P(\mu|L,L_1)\to P(\eta|\tilde{L},\tilde{L}_{1})=\Omega(\tilde{L},\tilde{L}_{1})^{-1}\left(\frac{\eta}{e^{\eta}-1}
\right)^{\tilde{L}}e^{\eta \tilde{L}_{1}},
\label{eq:Peta}
\end{equation}
where
\begin{equation}
\Omega(\tilde{L},\tilde{L}_{1})\equiv\int_{-\infty}^{\infty}\left(\frac{\eta}{e^{\eta}-1}
\right)^{\tilde{L}}e^{\eta \tilde{L}_{1}}\,d\eta.
\end{equation}
The limiting distribution~\eqref{eq:Peta} allows us to obtain $\eta$ for any dataset as the maximum-likelihood estimate. Differentiating
\begin{equation}
\log P(\eta|\tilde{L},\tilde{L}_{1})=\tilde{L}\left[\log\eta-\log(e^{\eta}-1)+\eta\frac{\tilde{L}_{1}}{\tilde{L}}\right]
-\log\Omega(\tilde{L},\tilde{L}_{1})
\end{equation}
with respect to $\eta$ leads to the equation
\begin{equation}
\frac{\tilde{L}_{1}}{\tilde{L}}=\frac{e^{\eta}}{e^{\eta}-1}-\frac{1}{\eta}=g(\eta).
\label{eq:fit_eta1}
\end{equation}
Comparing this equation to \eqref{eq:geta} provides the interpretation
\begin{equation}
\frac{\tilde{L}_{1}}{\tilde{L}}=t=\frac{s_{\text{max}}-\sigma}{s_{\text{max}}-s_{\text{min}}}.
\label{eq:fit_eta2}
\end{equation}

As in the discrete case, equation~\eqref{eq:fit_eta2} enables us to estimate the value of the parameter $\eta$ given the total cost per item $\sigma$ from a set of empirical data. There is an important difference though: we now need to set the scale of costs, namely the values of $s_{\rm max}$ and $s_{\rm min}$, using additional information on the dataset---a problem that did not arise in the discrete case because the first and last categories were fixed. Remember that $s_{\rm max}$ defines the largest possible cost that one can invest in one item, whereas $s_{\rm min}$ defines the least possible such cost. Once these parameters are known, $t$ is estimated as
\begin{equation}
t=\frac{s_{\text{max}}-\bar s}{s_{\text{max}}-s_{\text{min}}}, \qquad \bar s=\frac{1}{\tilde L}\sum_{i=1}^{\tilde L}s_i,
\end{equation}
where the $s_i$ are the costs associated to each of the items $i=1,\dots,{\tilde L}$, measured in the same units as $s_{\rm max}$ and $s_{\min}$.

For the confidence interval of the maximum-likelihood estimate of $\eta$ we need to introduce the function
\begin{equation}
\Phi_u(R)\equiv\int_0^u\left(\frac{\eta}{1-e^{-\eta}}\right)^{\tilde{L}}e^{-R\eta}\,d\eta.
\end{equation}
Then the $1-2\delta$ confidence interval for $\eta$, given $\tilde{L}$ and $\tilde{L}_{1}$, is obtained
through the cumulative distribution
\begin{equation}
\Gamma(u|\tilde{L},\tilde{L}_{1})=\int_{-\infty}^uP(\eta|\tilde{L},\tilde{L}_{1})\,d\eta=
\begin{cases}
\displaystyle \frac{\Phi_{\infty}(\tilde{L}_{1})-\Phi_{-u}(\tilde{L}_{1})}{\Phi_{\infty}(\tilde{L}_{1})
+\Phi_{\infty}(\tilde{L}-\tilde{L}_{1})}, & u<0, \\[5mm]
\displaystyle \frac{\Phi_{\infty}(\tilde{L}_{1})+\Phi_u(\tilde{L}-\tilde{L}_{1})}{\Phi_{\infty}(\tilde{L}_{1})
+\Phi_{\infty}(\tilde{L}-\tilde{L}_{1})}, & u\ge 0.
\end{cases}
\label{eq:Gamma}
\end{equation}
More precisely, the confidence interval $[\eta_-,\eta_+]$ is determined by solving the equations
$\Gamma(\eta_-|\tilde{L},\tilde{L}_{1})=\delta$, and $\Gamma(\eta_+|\tilde{L},\tilde{L}_{1})=1-\delta$ (see \hyperref[sec:methods]{Methods} for numerical details). In what follows we choose a $95\%$ confidence interval using $\delta = 0.025$.

\subsection{Mobile phones dataset}
\label{sec:mobile}

We have obtained the first dataset to analyse from Ref.~\citeonline{saramaki2014persistence} (actually, data were originally collected for another study\cite{roberts2011costs}). This dataset contains the phone activity of 24 individuals during 18 months. At the beginning of the study, all participants (12 males, 12 females, ages 17-19) were in their final year of secondary school, so that about six months later they transitioned into either university (18 of them) or labour market. The data from the phones (which were given for free to the participants along with 500 free monthly voice minutes and unlimited text messages) were complemented with three questionnaires, one at the beginning of the study, another one at the end of month 9, and a last one at the end of month 18. With this information, the authors were able to merge phone numbers that belonged to the same person, and, most importantly, to conclude that the number of calls was a reliable estimate of the emotional closeness of the relationships (see Ref.~\citeonline{saramaki2014persistence} for details).

In the original study, the communication patterns of the participants were analysed by dividing the dataset into three time intervals ($T_1$, $T_2$, and $T_3$) of six months each. For each time interval, the number of calls from each ego to each alter were counted and the alters were subsequently ranked based on this number. Then, the curve representing the fraction of calls as a function of the alters' ranks is used as a fingerprint of the ego's communication pattern. The main result of this study is that, even though the composition of personal networks varies considerably over time, these patterns are consistent across the different time windows. They named these patterns \emph{social signatures} and conjectured that they were likely a consequence of a constraint on the available resources (time and cognitive skills) necessary to manage relationships. 

In order to analyse these data we first aggregate them into the same time windows, so that we end up with a list (per time window) of pairs $(a_i,n_i)$ for each ego, where $a_i$ is a given alter and $n_i$ is the total number of calls made to that alter. As we explained in section~\ref{sec:bayesian_estimate}, prior to fitting the model we need to determine what $s_{\rm min}$ and $s_{\rm max}$ are for each participant (at each time window). To that end, we first select the minimum and the maximum number of calls each ego made every month to any alter. Then, $s_{\rm min}$ (respectively $s_{\rm max}$) for each time window is defined as the sum of the monthly minima (respectively maxima) along the six-month period. The rationale for these definitions is that these would have been the maximum and minimum number of calls to an alter, had this alter been the same all along the time window. 
Once $s_{\rm min}$ and $s_{\rm max}$ have been determined, we filter out any interaction below $s_{\rm min}$ (alters receiving fewer calls do not qualify as true relationships) and fit the model as explained in section~\ref{sec:bayesian_estimate}.

\begin{figure*}[!t]
\centering
\includegraphics[scale=0.70]{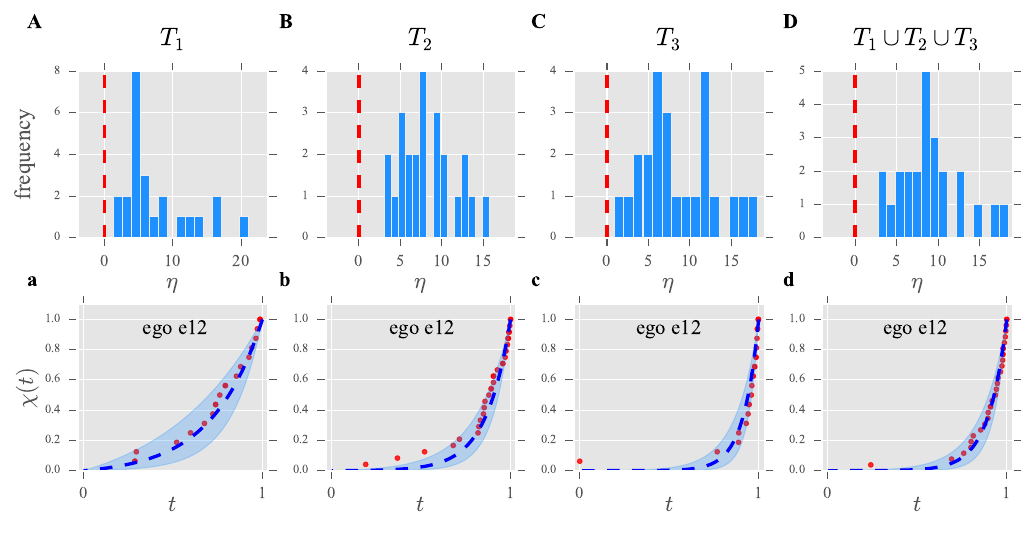}
\caption{\textbf{Summary of the results for the mobile phones dataset.}
Upper panels show the distributions of the parameter estimates for the different time windows (see section~\ref{sec:mobile} for details). The red, dashed lines mark the change from standard ($\eta>0$) to inverse ($\eta<0$) regimes.
\textbf{A,} Distribution of the parameter estimates for the first time window (months 1-6); ${\it mean}=7.52$, ${\it median}=5.32$, ${\it std}=5.07$.
\textbf{B,} Distribution of the parameter estimates for the second time window (months 6-12); ${\it mean}=8.32$, ${\it median}=8.00$, ${\it std}=3.31$.
\textbf{C,} Distribution of the parameter estimates for the second time window (months 12-18); ${\it mean}=8.48$, ${\it median}=7.00$, ${\it std}=4.56$.
\textbf{D,} Distribution of the parameter estimates for the full time window (months 1-18); ${\it mean}=9.07$, ${\it median}=8.75$, ${\it std}=3.92$.
Lower panels show the fittings for the same individual (ego ``e12'') at each of the time windows. Solid dots represent experimental data, blue dashed lines represent the graph of $\chi(t)$ in equation~\eqref{eq:chit} with the corresponding estimated parameter, and shaded regions show the 95\% confidence interval for that estimate (see section~\ref{sec:bayesian_estimate}).  
\textbf{a,} Example of fitting for an individual (``e12'') in the first time window. Estimated $\eta=3.55$, $95\%$ confidence interval $(1.82,5.77)$, $\tilde{L}=21$.
\textbf{b,} Example of fitting for an individual (``e12'') in the second time window. Estimated $\eta=7.38$, $95\%$ confidence interval $(5.18,10.34)$, $\tilde{L}=33$.
\textbf{c,} Example of fitting for an individual (``e12'') in the third time window. Estimated $\eta=11.79$, $95\%$ confidence interval $(7.87,17.69)$, $\tilde{L}=23$.
\textbf{d,} Example of fitting for an individual (``e12'') in the full time window. Estimated $\eta=9.77$, $95\%$ confidence interval $(6.83,13.95)$, $\tilde{L}=30$.
A comprehensive set of figures, including fittings for every subject at all time windows, is available in the Supplementary Information.}
\label{fig:phones}
\end{figure*}

Figure~\ref{fig:phones} summarises our results. As we can see in panels A-D, the distributions of the parameter estimates are centred around values consistent with the predicted $\eta\approx 6$ scaling (see section~\ref{sec:connection}). Additionally, the model is able to capture individual's nuances (panels a-d), and the fittings are, generally speaking, strikingly good (see Supplementary Information for a comprehensive set of figures, including fittings for every subject within every time window). Furthermore, we find a very high, significant correlation between the estimated parameter for each ego and the number of alters in his or her network ($\tilde{L}$). More precisely: $\eta^{T_1}\sim\tilde{L}^{T_1}$ ($r=0.84,p<10^{-6}$), $\eta^{T_2}\sim\tilde{L}^{T_2}$ ($r=0.52,p<10^{-3}$), $\eta^{T_3}\sim\tilde{L}^{T3}$ ($r=0.81,p<10^{-5}$) and $\eta^{T_1\cup T_2\cup T_3}\sim\tilde{L}^{T_1\cup T_2\cup T_3}$ ($r=0.83,p<10^{-6}$)---Pearson's $r$ coefficients, 2-tailed tests. This fact further endorses the claim that the amount of resource available to form relationships is a seemingly fixed quantity that individuals spread according to the maximum entropy principle.\cite{tamarit:2018}

Lastly, we analyse if the parameter $\eta$ may serve as a quantitative characterisation of the social signatures. In Ref.~\citeonline{saramaki2014persistence}, the authors used the Jensen-Shannon divergence\cite{lin1991divergence} (JSD) to measure the shape difference (distance) between signatures. Sticking to the notation in that reference, we will denote $d^{ij}_{ab}$ the JSD distance between the signature of ego $i$ in time $a$ and ego $j$ in time $b$. This measure was used to compute the variation between the signatures of the same ego ($i$) in consecutive time windows as $d^{ii}_{12}\equiv d^{\rm self}_{12}(i)$ and $d^{ii}_{23}\equiv d^{\rm self}_{23}(i)$. For comparison, the authors also computed the reference distances 
\begin{equation}
d^{\rm ref}_{22}(i)=\frac{1}{N_{\rm egos}-1}\sum_{j\ne i}d^{ij}_{22},
\> d^{\rm ref}_{33}(i)=\frac{1}{N_{\rm egos}-1}\sum_{j\ne i}d^{ij}_{33},
\end{equation}
and found that these reference distances were consistently higher than the ones between signatures of the same ego.\cite{saramaki2014persistence} 

\begin{figure*}[!t]
\centering
\includegraphics[scale=0.6]{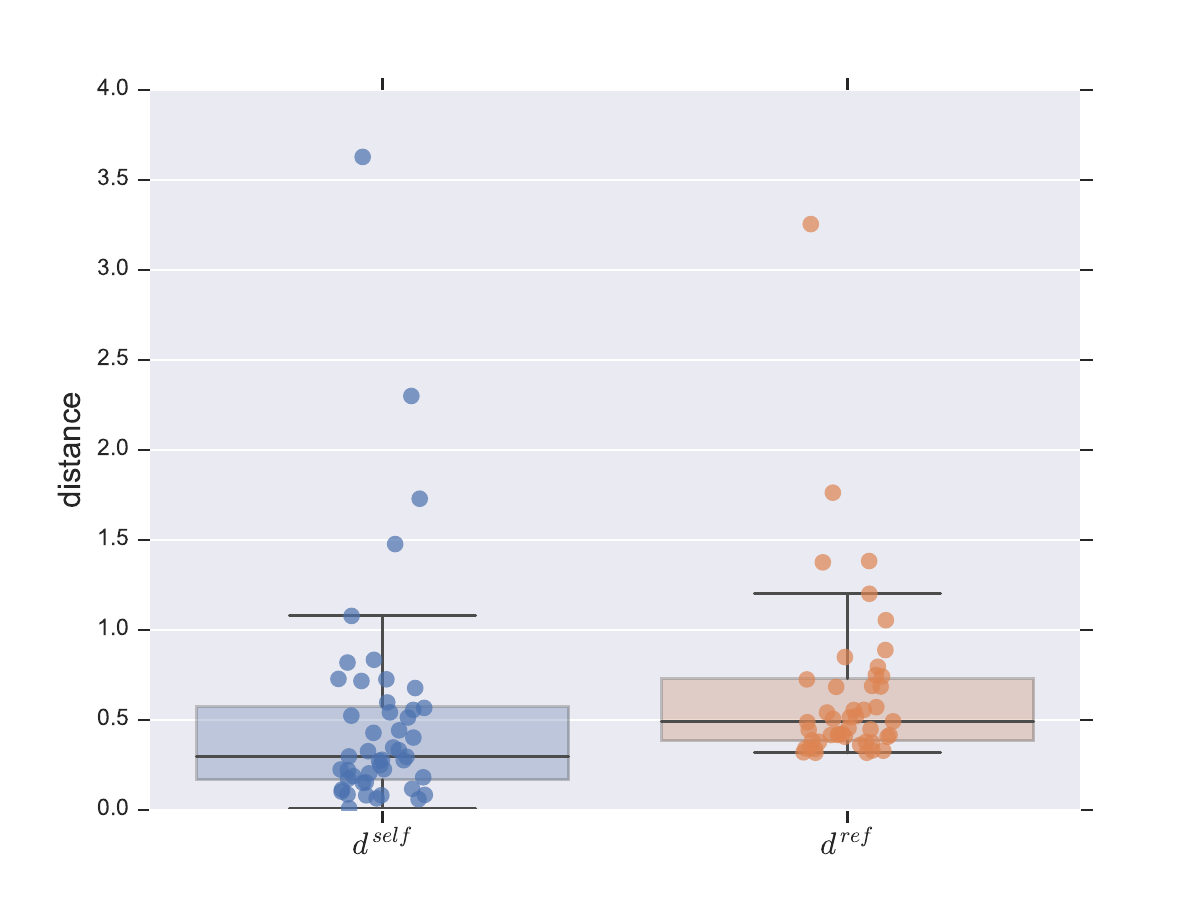}
\caption{\textbf{Evidence of the persistence of $\eta$ through time windows.} 
The boxplot to the left (blue) shows the distribution of distances between the parameter estimates for the same individual at consecutive time intervals ($d^{\rm self}$). The boxplot to the right (orange) shows the distribution of reference distances between the parameter estimate for each individual and the rest of the population ($d^{\rm ref}$). In both cases, the solid dots represent the empirical points---\emph{jittered} for a better visualisation. The distances in $d^{\rm ref}$ are consistently higher than those in $d^{\rm self}$, meaning that the individual's $\eta$ tends to be persistent across time intervals (see section~\ref{sec:mobile} for details).
}
\label{fig:persistence}
\end{figure*}

We perform a parallel analysis using the relative change between two different values of $\eta$ as a measure of the ``distance'' between them. That is, using the same notation, self-distances are obtained as \begin{equation}d^{\rm self}_{12}(i)\equiv \frac{|\eta_1^{i}-\eta_2^{i}|}{|\eta_1^{i}|},\> d^{\rm self}_{23}(i)\equiv\frac{|\eta_2^{i}-\eta_3^{i}|}{|\eta_2^{i}|},
\end{equation}
whereas the reference distances are given by
\begin{equation}
    d^{\rm ref}_{22}(i)=\frac{1}{N_{\rm egos}-1}\sum_{j\ne i}\frac{|\eta_2^{i}-\eta_2^{j}|}{|\eta_2^{i}|}, \> d^{\rm ref}_{33}(i)=\frac{1}{N_{\rm egos}-1}\sum_{j\neq i}\frac{|\eta_3^{i}-\eta_3^{j}|}{|\eta_3^{i}|}.
\end{equation} 
We then create a distribution of self-distances $d^{\rm self}=\bigcup_{i}^{N_{\rm egos}}\{d^{\rm self}_{12}(i),d^{\rm self}_{23}(i)\}$ as well as another one of reference distances $d^{\rm ref}=\bigcup_{i}^{N_{\rm egos}}\{d^{\rm ref}_{22}(i),d^{\rm ref}_{33}(i)\}$ (a total of 48 points per distribution).
In Fig.~\ref{fig:persistence} we show the resulting distributions of self- ($d^{\rm self}$) and reference distances ($d^{\rm ref}$). The distribution $d^{\rm ref}$ is again consistently higher than that of $d^{\rm self}$---which is confirmed by a Mann-Whitney U test yielding $p<10^{-3}$ (two-sided). Therefore, the different egos tend to have a persistent value of $\eta$ just like they have a persistent social signature. Given that the central premise of our model is that the resources available to create relationships are limited (see section~\ref{sec:model}), this result reinforces the conjecture\cite{saramaki2014persistence} that the existence of social signatures is a consequence of this very constraint.

\subsection{Face-to-face contacts dataset}
\label{sec:sociopatterns}

In this section we analyse data from face-to-face interactions\cite{Isella2011crowd} that took place during a scientific conference in Turin, Italy, in 2009 (see \hyperref[sec:methods]{Methods}). The data were collected using proximity sensors that voluntary participants ($n=111$, about $75\%$ of the attendees) had embedded in their conference badges. The sensors recorded interactions over intervals of $20s$ when two or more participants were facing each other at less than about $1.5-2m$ (see Refs.~\citeonline{alani2009live,van2010live,cattuto2010dynamics,Isella2011crowd} for technical details). With this information, we can build the network of interactions for each participants using the time spent together as a proxy of the intensity of the implied relationships.

The high temporal resolution of the data permits us to characterise the values of $s_{\rm min}$ and $s_{\rm max}$ in several ways. One natural option is to aggregate the data over one day,\cite{Isella2011crowd} and use a similar rule to the one we applied in section~\ref{sec:mobile}---that is, use the sum of the maximum time spent with any alter on each day as $s_{\rm max}$, and the sum of the minima as $s_{\rm min}$. However, during a conference, many different time restrictions may apply to the attendees, such as having an agenda of presentations to attend or deliver. As a consequence, the aforementioned heuristic may not apply here, since it is very likely the case that it was not entirely up to the participants with whom to spend their time at a given moment. Furthermore, we do not have any information on the interactions with the $25\%$ of individuals who were at the venue but chose not to participate. These facts impose clear limitations to the conclusions we can draw from applying our model, and they are hardly avoidable. Therefore, we adopt a rather cautious position and do not aggregate the data on daily time windows. Instead, we simply take $s_{\rm max}$ as the maximum time spent (and recorded) with one alter during the whole conference, and $s_{\rm min}$ as the minimum one. Additionally, we exclude all participants who had fewer than five alters in their networks, ending up with a total of $95$ valid cases.

\begin{figure*}[!t]
\centering
\includegraphics[scale=1]{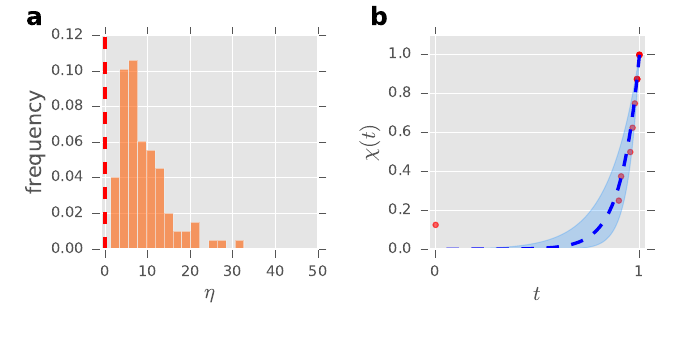} 
\caption{\textbf{Summary of the results for the face-to-face contacts dataset.}
\textbf{a,} Distribution of the parameter estimates for the face-to-face contacts dataset ($n=95$). The red, dashed line marks the change of regime ($\eta=0$); ${\it mean}=9.08$, ${\it median}=7.35$, ${\it mode}=5.54$, ${\it std}=5.86$.
\textbf{b,} Representative fitting for an individual in the face-to-face contacts dataset (chosen at random from those with a strictly positive $95\%$ confidence interval). Solid dots represent experimental data, blue dashed lines represent the graph of equation~\eqref{eq:chit} with the corresponding estimated parameter, and shaded regions show the 95\% confidence interval for that estimate (see section~\ref{sec:bayesian_estimate}). Estimated $\eta=11.12$, $95\%$ confidence interval $(6.74,18.34)$, $\tilde{L}=15$. See Fig.~S6 in the Supplementary Information for a sample of 24 other fittings chosen randomly from the entire population.}
\label{fig:sociopatterns}
\end{figure*}

Our results (Fig.~\ref{fig:sociopatterns}a) show a long-tailed distribution for the parameter estimates with a clear peak close, once again, to the predicted $\eta\approx 6$, which suggests that the overall behaviour of the contact patterns seems to agree with our model.
However, even though some fittings are quite good (see Fig.~\ref{fig:sociopatterns}b), overall they are not as good as those of the mobile phones dataset (see Fig.~S6 in the  Supplementary Information).
For comparison, we also carried out the analysis using the same approach as in section~\ref{sec:mobile} to set $s_{\rm min}$ and $s_{\rm max}$. Figure~S5 in the Supplementary Information collects the corresponding results, showing individual fits that are slightly worse and distributions of the parameter estimates centred around a higher value ($\eta\approx 14$). 
It has to be taken into account that, 
as explained above, these data are inherently noisy and assessing the intensity of the relationships (or even merely of the interactions) based solely on time spent together during a conference can be misleading. Ideally, we would need this type of data but from individuals in their daily lives, so that the interactions recorded would better correspond to decisions of the individual. Nevertheless, even with the mentioned limitations, the model is still capable of capturing the patterns of face-to-face interactions to some extent.

\subsection{Facebook dataset}

If we compare the results from sections~\ref{sec:mobile} and \ref{sec:sociopatterns} (Figs.~\ref{fig:phones}a and \ref{fig:sociopatterns}a) we can appreciate how, as the sample size increases, the distribution of the parameter estimates seems to smooth around a well-defined central value $\eta\approx 6$. If that were the case, it would be a clear indication that the parameter of the model is indeed capturing a real feature of the way individuals manage relationships.  
To further explore this possibility, we analyse a larger dataset of interactions in Facebook.\cite{arnaboldi2012analysis} This dataset was obtained using a crawler on April 2008 and comprises data on roughly 3 million Facebook users and 23 million edges. Importantly, it also contains the number of interactions (photo comments or Wall posts) between users. The data is divided into four different time windows (referred to the time of the crawl): last month, last six months, last year and all---which contains all the interactions among the users since they established their links.\cite{arnaboldi2012analysis}

To analyse the structure of the personal networks in Facebook, the authors of that study filtered the data to retain only active, relevant users from which the relative frequency of contact with all his or her alters could be adequately assessed (see Ref.~\citeonline{arnaboldi2012analysis} for details). The resulting dataset contains about $90,000$ users and $4.5$ million links. Applying two different clustering techniques, $k$-means\cite{wang2011ckmeans} and DBSCAN,\cite{ester1996density} they found that the structure of personal networks of Facebook users consists of a set of $4$ concentric, inclusive circles according to the intensity of their links, and that the sizes of these circles exhibited a more or less constant scaling ratio close to 3---thus, resembling what is found in offline social networks.\cite{dunbar2015structure}

\begin{figure*}[!t]
\centering
\includegraphics[scale=1]{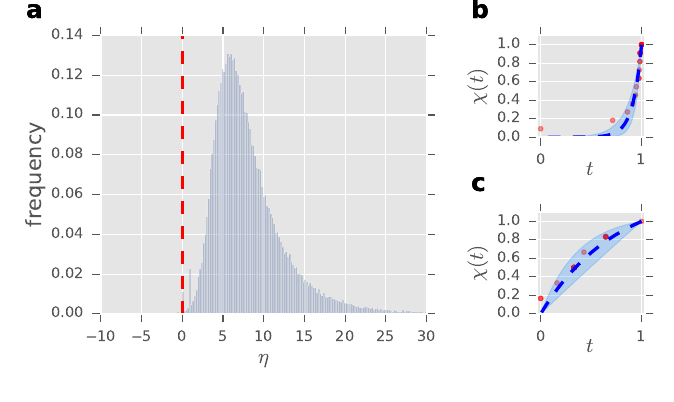} 
\caption{\textbf{Summary of the results for the Facebook dataset.}
\textbf{a,} Distribution of the parameter estimates for the Facebook dataset ($n=98,258$). The red, dashed line marks the change of regime $\eta = 0$; ${\it mean}=8.25$, ${\it median}=7.17$, $mode=5.48$, ${\it std}=4.91$
\textbf{b,} Representative fitting for an individual exhibiting the standard regime (chosen at random from those with a strictly positive $95\%$ confidence interval). Solid dots represent experimental data, blue dashed lines represent the graph of equation \eqref{eq:chit} with the corresponding estimated parameter, and shaded regions show the $95\%$ confidence interval for that estimate (see section \ref{sec:bayesian_estimate}). Estimated $\eta=11.64$, $95\%$ confidence interval $(7.62,17.79)$, $\tilde{L}=21$.
\textbf{c,} Example of fitting for an individual exhibiting the inverse regime (chosen at random from those with a strictly negative $95\%$ confidence interval). Solid dots, blue dashed lines, and shaded regions have the same interpretation as in \textbf{b}. Estimated $\eta=-1.34$, $95\%$ confidence interval $(-2.71,-0.08)$, $\tilde{L}=30$. See Fig.~S7 in the Supplementary Information for a sample of 24 other fittings chosen randomly from the entire population.
}
\label{fig:facebook}
\end{figure*}

Since clustering algorithms find an optimal partition of personal networks into four circles with a scaling of approximately 3, our model should yield a distribution of parameters centred around $\eta\approx 6$. In this case, for each individual $s_{\rm max}$ is simply given by his or her most intense interaction, and $s_{\rm min}$ by the least intense one---with this decision, we can use the original dataset without any further pre-processing. Figure~\ref{fig:facebook} confirms our hypothesis, showing a smooth distribution with ${\it mean}=8.25$, ${\it median}=7.17$, and $mode=5.48$. Interestingly, the size of this sample allows us to find, for the first time, individuals exhibiting an inverse regime ($\eta<0$). Specifically, we find 256 users, about $0.3\%$ of the population, exhibiting this type of structure---to be precise, only for $7$ of them ($0.007\%$) the $95\%$ confidence interval does not include the zero. In Figs.~\ref{fig:facebook}b and \ref{fig:facebook}c we show representative fits of individuals in the standard and the inverse regime, respectively. Let us remark that not only does our model capture the typical structure of personal networks,\cite{arnaboldi2012analysis} but it also unveils that the inverse regime\cite{tamarit:2018} can also be found in digital communications---in spite that this is the last environment one would expect to find it because of the usual inflation of contacts it favours.

\section{Discussion}

In this paper we have presented an extension of the discrete model of costly allocation of resources introduced elsewhere,\cite{tamarit:2018} which treats the cost as a continuous variable. While our approach allows us to deal with any such problem of resource allocation, we have applied it to case of the structure of personal networks when the intensity of emotional links is given by a continuous magnitude (time spent with the alter, number of phone calls, messages exchanged, etc.) which cannot be naturally classified in categories or layers of intensity. We have found that the behaviour of this continuous model is qualitatively identical to that of its discrete counterpart. Remarkably, our experimental results show that the estimates of the new parameter characterising the distribution of links ($\eta\approx 6$) are consistent with the scaling relation between circles typically observed in discrete settings ($\mu\approx 1$ or $e^{\mu}\approx 3$). Consequently, one may wonder whether the organisation of personal networks has a discrete (as empirical evidence has suggested so far) or continuous nature.

Given the abundant empirical evidence for the existence of discrete layers, we are inclined to think that the discretisation might be real---if only because of the natural human tendency to classify and the inherent difficulty to deal with the continuum. However, this discretisation will hardly be perfect and may be subject to fluctuations. Moreover, even if the (psychological) organisation of the networks were perfectly discrete, it would be difficult for all people within the same layer to receive precisely the same attention (number of calls, contact time, and so on) at all times, which would cause continuous fluctuations. Let us emphasise that under no circumstances are both results incompatible, since our (continuous) model does not assume at any time that the distribution of intensities is continuous, but only that it can be so measured. The model we have developed simply allows us to manage this type of data without having to make \emph{ad hoc} assumptions on the number of layers. Importantly, the principles underlying both types of structures are indeed the same, namely that relationships are costly in terms of (cognitive) resources and that the we have a limited amount of these resources to devote to them.

The use of the continuum approach we have introduced here has its own drawbacks. Dispensing with the layers/circles allowed us to find a parameter that characterises the scaling of the distribution of resources valid in any situation, but the price to pay is that the scale in which the intensity of the relations is measured (i.e., $s_{\rm min}$ and $s_{\rm max}$) has to be inferred from additional information on the problem. This creates a further challenge when fitting the data, and decisions have to be made based on plausible reasons---but there might be other possibilities. This might well be one of the reasons why the individual fittings seem to be somewhat worse than the ones obtained with the discrete model,\cite{tamarit:2018} and it is an issue that deserves further attention. 

On the other hand, it is important to realize that one of the assumptions of the model is that the effort devoted to relationships is a perfect indicator of their intensity. This must be compared with the different types of information with which we have measured these efforts (number of calls, face-to-face contact, and number of messages exchanged), which are nothing more than proxies for that effort. In particular, although contacts can be maintained using different means (phone calls, personal meetings, Facebook, etc.), in our analyses we are focusing only on one of them. Including all the data of contacts among people through any means should improve the results. Another relevant issue is that, more likely than not, all communications are not equally intense, even if their duration is the same, which is a significant source of noise for our model. In any case, given the simplicity of the model and the particularities of the data, the fits are remarkably good. Furthermore, the aggregate distribution of the parameter estimates (which might compensate for individual errors) exhibits a clear shape centred around the expected value of $\eta\approx 6$, a remarkable result in itself that makes it clear that our relationships exhibit the signature of a resource allocation problem. 


\section*{Methods}
\label{sec:methods}

All numerical analyses are carried out in \texttt{Python} with the packages \texttt{scipy.optimize} and \texttt{scipy.integrate}. The documentation of these packages can be found in \url{https://docs.scipy.org/doc/scipy-0.14.0/reference/generated/scipy.optimize.fsolve.html}. 

To compute the integrals in equation ~\eqref{eq:Gamma} for finite values of $u$ we use the function \texttt{quad} (\texttt{Python}). For $u\to\infty$ we evaluate
them using a Gauss-Laguerre quadrature with $150$ points. Overflows due to exponentiation are avoided by evaluating the logarithm of the integrand, and the singularity at $\eta = 0$ is avoided by Taylor expanding $e^{-\eta r}$ and $e^{-\eta}$ up to third order. Likewise, the singularity at $\eta=0$ of \eqref{eq:fit_eta1} is avoided by using the Taylor expansion $\chi_k \approx k/r + (k/2r)(e^{\eta}-1)(k-r)$ for $|e^{\eta}-1| \leq 10^{-6}$. The extremes of the confidence interval $[\eta_{-},\eta_{+}]$ and the equation ~\eqref{eq:fit_eta1} are solved using the function \texttt{fsolve} with tolerance $10^{-6}$. The code used for these analyses is publicly available.\cite{tamarit:2021}

Data for the analysis of section~\ref{sec:sociopatterns} has been downloaded from the SocioPatterns webpage\cite{sociopatterns:2009} (last accessed 24 January 2019). They register face-to-face interactions that took place during the scientific conference ``Hypertext 2009: 20th ACM Conference on Hypertext and Hypermedia'' (\url{http://www.ht2009.org/}), held in Turin, Italy, between June 29th and July 1st in 2009.

The Facebook dataset used to be available, upon request, at \url{http://current.cs.ucsb.edu/socialnets/} under the name ``Anonymous regional network A''. However, as of April 24, 2021, it seems that the web is no longer available. We obtained the data thanks to Prof.~Ben Zhao's kindness.

\bibliography{refs}

\section*{Acknowledgements}

We are very thankful to Prof.~Ben Zhao for granting us access to his Facebook dataset, and to Dr.~Valerio Arnaboldi for sharing with us the final, curated dataset just as they used it in their paper. This research has been funded by the Spanish Ministerio de Ciencia, Innovaci\'on y Uni\-ver\-si\-da\-des-FEDER funds of the European Union support, under project BASIC (PGC2018-098186-B-I00).

\section*{Author contributions statement}

All three authors (I.T., A.S., and J.C.) conceived the research, analysed and discussed the results and wrote the manuscript. I.T. generated the fits to the datasets and the numerical work. I.T. and J.C. developed the mathematical model.

\section*{Additional information}

To include, in this order: \textbf{Accession codes} (where applicable); \textbf{Competing interests} (mandatory statement). 

\subsection*{Accession codes}

\subsection*{Competing interests}

The authors declare no competing interests.
\end{document}